# Medicinal Plants Database and Three Dimensional Structure of the Chemical Compounds from Medicinal Plants in Indonesia

**Arry Yanuar[1*], Abdul Mun'im[1], Akma Bertha Aprima Lagho[1], Rezi Riadhi Syahdi[1], Marjuqi Rahmat[2], and Heru Suhartanto[2]**

**[1] Department of Pharmacy, Faculty of Mathematics and Natural Sciences, University of Indonesia**
**Depok, 16424, Indonesia**

**[2] Faculty of Computer Sciences, University of Indonesia**
**Depok, 16424, Indonesia**

**Abstract**

During this era of new drug designing, medicinal plants had become a very interesting object of further research. Pharmacology screening of active compound of medicinal plants would be time consuming and costly. Molecular docking is one of the *in silico* method which is more efficient compare to *in vitro* or *in vivo* method for its capability of finding the active compound in medicinal plants. In this method, three-dimensional structure becomes very important in the molecular docking methods, so we need a database that provides information on three-dimensional structures of chemical compounds from medicinal plants in Indonesia. Therefore, this study will prepare a database which provides information of the three dimensional structures of chemical compounds of medicinal plants. The database will be prepared by using MySQL format and is designed to be placed in **http://herbaldb.farmasi.ui.ac.id** website so that eventually this database can be accessed quickly and easily by users via the Internet.

***Keywords:*** *Indonesia's Medicinal Plants, Natural Compounds Database, Three Dimensional Structure.*

## 1. Introduction

Indonesia as a mega-biodiversity center has a second position in the world after Brazil. Indonesia become the first position, if marine biota is also included. In our Earth, which lived about 40,000 species of plants, of which 30,000 species live in the Indonesian archipelago. Among the 30,000 plant species that live in the Indonesian archipelago known to at least 9600 species of plants have a pharmacological activity [1]. In an era of new drug design, these plants become the basic material to study further. Screening of pharmacological activity of active ingredients in medicinal plants is an expensive process, requiring energy, qualified human resources and require a long time if done in laboratory experiments using experimental animals [2]. Attractive chance today is the use of computers as tools in drug development. Exponentially increase computing capabilities provide opportunities to develop simulations and calculations in drug design. The method used in drug design can be a structure-based drug design and ligand-based drug design [3]. In the field of structure-based drug design, molecular docking is a commonly used method. Molecular docking is a method which used to predict an intermolecular complex between the drug molecule with its target protein. When performing molecular docking, a set of data which contains information on the ligand or drug to be docked and protein targets to be used are needed. The information required in this process include three-dimensional structure of the ligand and target protein [4].

Until recently, there is no database of medicinal plants and its natural compounds in Indonesia, which presents data interactively and comes with three-dimensional structure of chemical compounds. The aim of this research is to prepare medicinal plants database and three dimensional structure of the active compounds from medicinal plants in Indonesia.

## 2. Design and Methods

This research is done using combination of literature study, simulation and molecular modeling. The presentation of the results as a monographs are reported descriptively through online database prototype. The experimental design is made as follows:

### 2.1 Screening and preparing the list of Indonesian medicinal plants

This study begins with gathering information about the collection of medicinal plants in Indonesia. Searching of medicinal plants in Indonesia obtained by literature from scientific journals, books and websites. This search is then continued with the selection of plants used as medicine in Indonesia. Various sources of classic and official books are used as a database raw data [5-12].

## 2.2 Searching and drawing of chemical compound

From medicinal plants found in the database table and then collected data on chemical compounds found in the medicinal plants. If the chemical compounds found contained in it, then extended the search to find a two-dimensional structure of these chemical compounds. Two-dimensional structure of the search query is performed on the chemical compound database Pubchem [13] or KNApSAcK [14]. Two-dimensional structure of the search should be available in specific file formats. In KNApSAcK metabolite database, available file format is gif file format to be converted into a file format acceptable to the program used. Conversion of file formats was done by drawing two-dimensional structures of chemical compounds of medicinal plants using Symyx Draw program. The 2D structures were then stored in a file .mol format [15].

## 2.3 Generating the 3D structure

Three-dimensional structure is formed using VEGAZZ [16], a compound modification program. File format .mol (storage results from Symyx Draw program) or file format .sdf which have been downloaded from PubChem (the conversion results with the program PyMOL [17]) are then included in the program VEGAZZ computer which will then be processed to form a new three-dimensional molecule. Formation of three-dimensional structure was done through several stages. The first stage is to open two-dimensional structure that is stored in a file format .mol. Two-dimensional structures that have emerged in Vega ZZ view, will then be turned into three-dimensional with the run command run scripts. Run this script and then bring up the next dialog box choose the command '2 D to 3D 'which means the two-dimensional display will be turned into a three-dimensional. Having formed three-dimensional structure, save the file with the .mol format. The .mol format was then converted into .mol2 format using the OpenBabel program [18].

## 2.4 Preparation of medicinal plants database.

Medicinal plant database was prepared using MySQL format. Designing databases using raw data (file. XLS) which has been established the previous step and armed with the requirements gathering process, then from there made it an entity relationship. The following entity relationship diagram illustrates the connectedness between the data to be stored in a database.

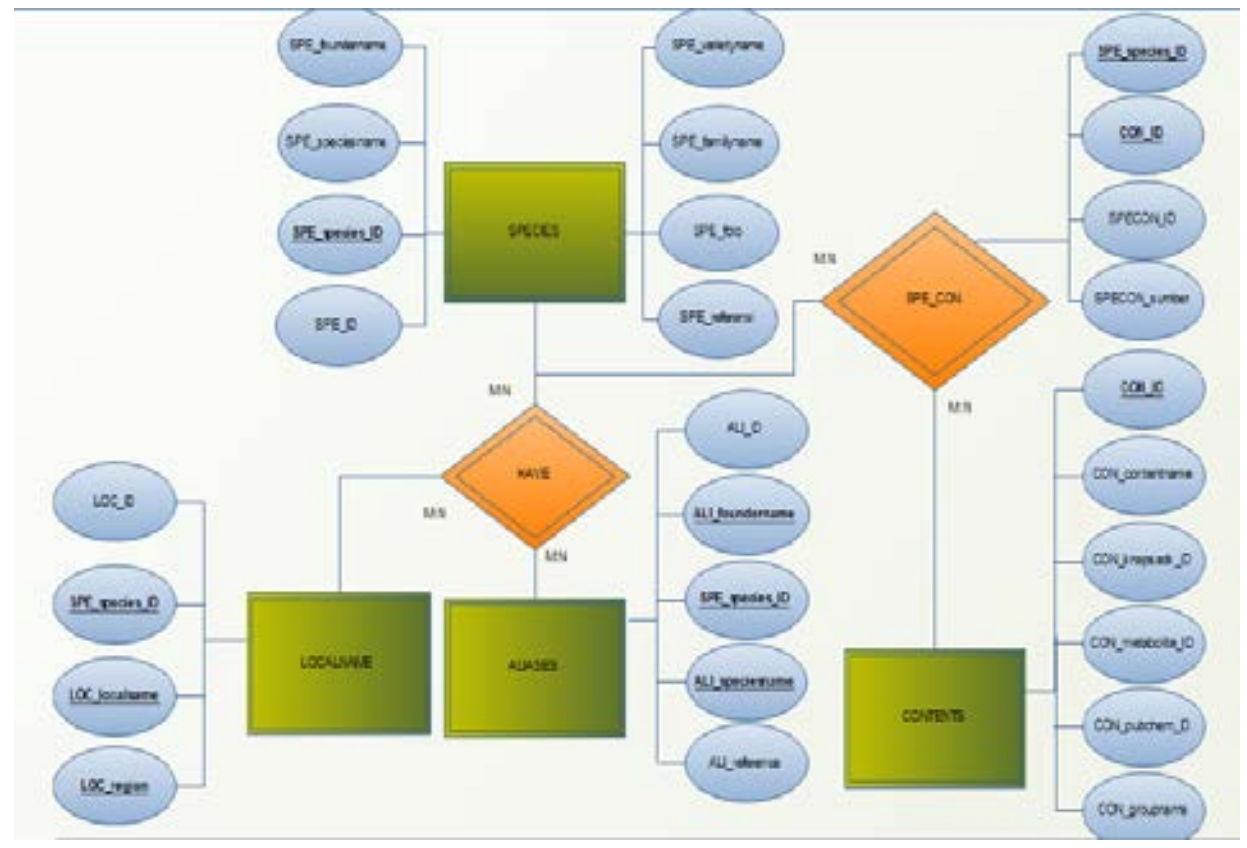

Fig 1. Entity Relationship Diagram

## 2.5 Creating a prototype web database

Prototype of web database of medicinal plants and three-dimensional structures of chemical compounds was created by PHP. The user interface is created are the homepage, the page data and details of species, compound data pages, profile pages, admin pages, file upload pages, user pages, search results pages, language settings and FAQs page. The authorized users are divided into the categories: administrator, expert and contributor that have a role to maintain the database via login session. The common users could also use this database on the website without login session, but they will have limited access.

# 3. Results and Discussion

To date data has been collected as many as 3825 species records, with various local names as many as 16,244 records, and content as many as 6776 record recorded in the (species-contents) interaction as many as 12,980 records. All of this data is collected from the various literature and noted the source. Web database prototype of medicinal plants and three-dimensional structures of chemical compounds have been created with PHP. Website address of Indonesian Medicinal Plant Database is located at URL http://herbaldb.farmasi.ui.ac.id. Initially, a total of 1412 three-dimensional structure of chemical compounds from medicinal plants of Indonesia embedded in the system. The entire database system runs well and database needs to be verified by the expert users.

The nature of open system allows wide use of plant medicine for Indonesia database by both the general public and scientists or other stakeholders from both industry sectors Herbs / Jamu, government, or university. The nature of “Open Systems”, a search algorithm and interactive data presentation allows the emergence of knowledge systems. Quality assurance and validity of the database content can be maintained by the control from the existing users and the verification from an expert. Thus the potential for growth and wider utilization becomes very open. Global utilization is also wide open with the features of English.

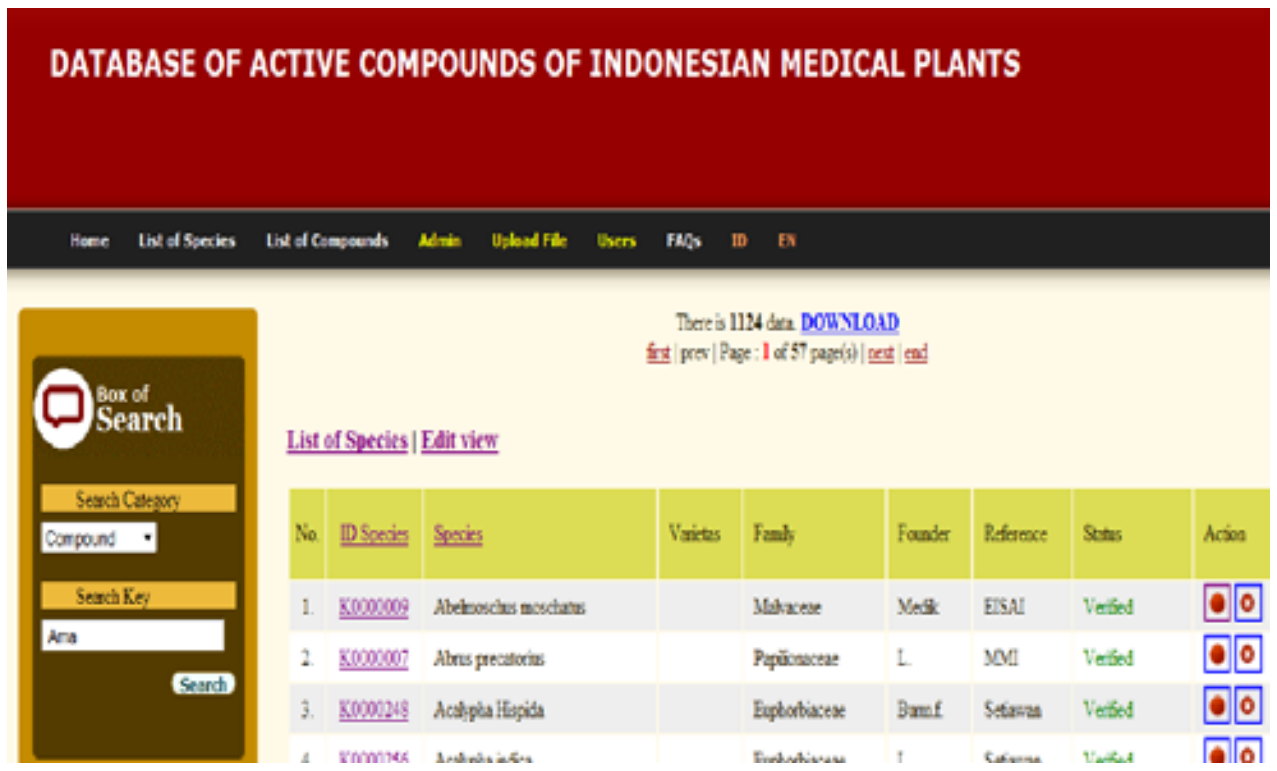

Fig 2. Web database snapshot of list of species

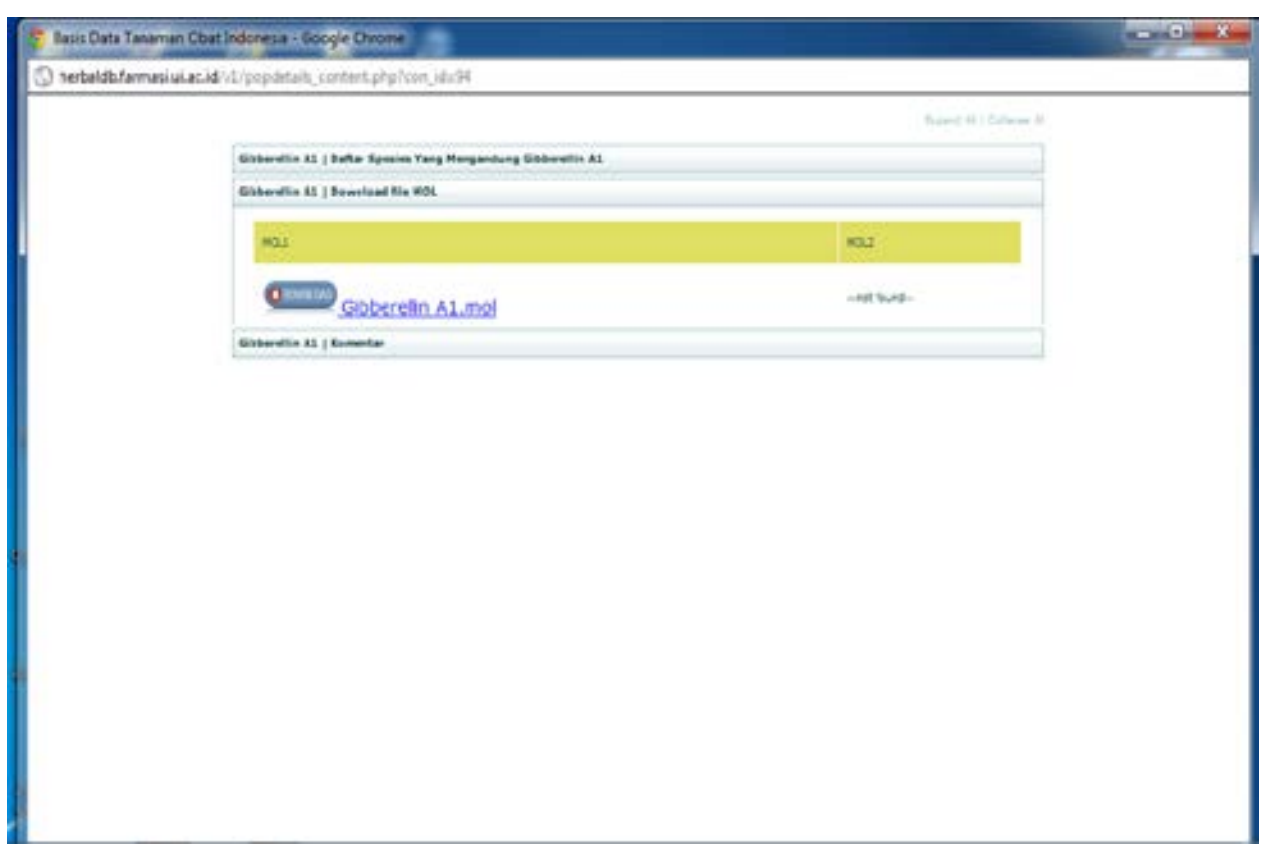

Fig. 3 Snapshot of 3D chemical structure download page.

## 4. Conclusions

The prototype of the web-based interactive Indonesian Medicinal Plant Database System has been created using PHP and MySQL relational database. Database contents can be accessed through a search algorithm using the species name, alias names or synonyms, local or regional names, compound name, and usage keywords. Data species are equipped with photographs that can be uploaded by the authorized user and the name of the chemical compound has three-dimensional data structure in the format .mol and .mol2 that can be uploaded by authorized users and downloaded by all users. The features of the contents are made as "open system" those can be modified, added or deleted by authorized users. The uniqueness of this system that is also seen as an open feedback system is a comment feature for species name and compound name of each entry, which allows for discussion between users and the objections of an entry. Authorized users are registered as administrator, expert or contributor, which each have a role in maintaining the content in the database. Indonesian medicinal plants database has a record with default: not verified. The task to verify the data in the database is performed by the expert users. This work is an early version of the overall road map website of Indonesian medicinal plants database and three-dimensional structure of chemical compounds. Website of Indonesian Medicinal Plant Database is located at **http://herbaldb.farmasi.ui.ac.id**.

### Acknowledgments

This research is financially supported by Riset Unggulan Universitas Indonesia (RUUI), 2010. A.Y. Author thanks to Prof. Shigehiko Kanaya, PhD from Nara Institute of Science and Technology (NAIST), Japan for providing data from KNApSAcK metabolite database.

## References

[1] Keputusan Menteri Kesehatan Republik Indonesia No : 381/MenKes/SK/III/2007, Menteri Kesehatan RI, Jakarta, 2007
[2] Jayaram, B., SCFBIO: What is drug design? http://www.scfbio-iitd.res.in/tutorial/drugdiscovery.htm,2011
[3] Hawkins, P., & Skillman, G., Ligand-based design workflow, http://images.apple.com/science/pdf/ligandbased_design_workflow.pdf, 2006
[4] Abraham, D.J., (ed). Burger's Medicinal Chemistry and Drug Discovery, Volume 1: Drug Discovery, 6th ed. Willey Interscience, 2003
[5] Heyne, K., De Nuttinge Planten van Indonesie, 3ed, Wageningen, H. Veenman & Zonen, 1950
[6] Departemen Kesehatan Republik Indonesia, Materia Medika Indonesia. Vol I. Jakarta: Departemen Kesehatan Republik Indonesia. 1977
[7] Departemen Kesehatan Republik Indonesia, Materia Medika Indonesia, Vol II. Jakarta: Departemen Kesehatan Republik Indonesia. 1978
[8] Departemen Kesehatan Republik Indonesia, Materia Medika Indonesia, Vol III. Jakarta : Departemen Kesehatan Republik Indonesia. 1979
[9] Departemen Kesehatan Republik Indonesia, Materia Medika Indonesia, Vol IV. Jakarta : Departemen Kesehatan Republik Indonesia. 1980
[10] Departemen Kesehatan Republik Indonesia, Materia Medika Indonesia, Vol V. Jakarta : Departemen Kesehatan Republik Indonesia. 1989
[11] Departemen Kesehatan Republik Indonesia, Materia Medika Indonesia, Vol VI. Jakarta : Departemen Kesehatan Republik Indonesia. 1995
[12] Medicinal Herb Index in Indonesia 2nd edition. Jakarta: PT. Eisai Indonesia, 1995
[13] Bolton, E., Wang, Y., Thiessen, P. A., and Bryant, S.H., PubChem: Integrated Platform of Small Molecules and

Biological Activities. Chapter 12 IN Annual Reports in Computational Chemistry, Volume 4, American Chemical Society, Washington, DC, 2008 , 217-241, DOI:10.1016/S1574-1400(08)00012-1

[14] Shinbo, Y., Nakamura, Y., Altaf-Ul-Amin, M., Asahi, H., Kurokawa, K., Arita, M., Saito, K., Ohta, D., Shibata, D. and Kanaya, S., KNApSAcK: A Comprehensive Species-Metabolite Relationship Database, Plant Metabolomics: in Biotechnology in Agriculture and Forestry, 2006, (57)165-181, DOI: 10.1007/3-540-29782-0_13

[15] Symyx Draw-An introductory guide, http://bbruner.org/obc/symyx.htm, 2011

[16] Pedretti, A., Villa. L., and Vistoli, G., Vega—an open platform to develop chemo-bio-informatics applications, using plug-in architecture and script programming. J. Comput. Aided. Mol. Des., 2004, 18(3) 167-173, DOI: 10.1023/B:JCAM.0000035186.90683.f2

[17] Delano, W., Pymol user's guide. Delano Scientific LLC.: http://pymol.sourceforge.net/newman/userman, 2004

[18] Guha, R., Howard, M. T., Hutchison, G. R., Murray-Rust, P., Rzepa, H., Steinbeck, C., Wegner, J. K., and Willighagen, E. L., "The Blue Obelisk -- Interoperability in Chemical Informatics." J. Chem. Inf. Model. (2006) 46(3) 991-998. DOI:10.1021/ci050400b

**Arry Yanuar** is an assistant Professor at Department of Pharmacy, Universitas Indonesia. He has been with Department of Pharmacy since 1990. He graduated from undergraduate program Department of Pharmacy, University of Indonesia in 1990. He also holds Pharmacist Profession certificate in 1991. In 1997, he finished his Master Program from Faculty of Pharmacy, Gadjah Mada University. He holds PhD in 2006 from Nara Institute of Science and Technology (NAIST), Japan, from Structural Biology/protein crystallography laboratory. In 1999-2003 he worked as pharmacy expert in ISO certification for pharmacy industries at Llyod Register Quality Assurance. In 2002, he visited National Institute of Health (NIH), Bethesda, USA. He won several research grants and published some paper at international journals and conferences.

**Abdul Mun'im** is an assistant Professor at Department of Pharmacy, Universitas Indonesia. He has been with Department of Pharmacy since 1990. His field of research is in natural product chemistry or phytochemistry.

**Akma Bertha Aprima Lagho** hold BSc from Department of Pharmacy, University of Indonesia in 2010.

**Rezi Riadhi Syahdi** is a Master Student from Department of Pharmacy, University of Indonesia. He hold BSc in Pharmacy in 2010.

**Marjuqi Rahmat** hold BSc from Faculty of Computer Sciences, University of Indonesia in 2010.

**Heru Suhartanto** is a Professor in Faculty of Computer Science, Universitas Indonesia (Fasilkom UI). He has been with Fasilkom UI since 1986. Previously he held some positions such as Post doctoral fellow at Advanced Computational Modelling Centre, the University of Queensland, Australia in 1998 – 2000; two periods vice Dean for General Affair at Fasilkom UI since 2000. He graduated from undergraduate study at Department of Mathematics, UI in 1986. He holds Master of Science, from Department of Computer Science, The University of Toronto, Canada since 1990. He also holds Ph.D in Parallel Computing from Department of Mathematics, The University of Queensland since 1998. His main research interests are Numerical, Parallel, Cloud and Grid computing. He is also a member of reviewer of several referred international journal such as journal of Computational and Applied Mathematics, International Journal of Computer Mathematics, and Journal of Universal Computer Science. Furthermore, he has supervised some Master and PhD students; he has won some research grants; holds several software copyrights; published a number of books in Indonesian and international papers in proceeding and journal. He is also member of IEEE and ACM.